\pdfoutput=1

\documentclass[preprint,12pt]{elsarticle}

\usepackage{graphicx}  
\usepackage{graphics}
\usepackage{dcolumn}   
\usepackage{bm}        
\usepackage{amssymb}   
\usepackage{color}
\usepackage{amsmath}
\usepackage{float}

\usepackage[font=scriptsize]{caption}

\journal{Physica A}

\begin{document}

\begin{frontmatter}



\title{The network asymmetry caused by the degree correlation and its effect on the bimodality in control}

\author[label1]{Xiaoyao Yu}
\author[label1]{Yongqing Liang}
\author[label1]{Xiaomeng Wang}
\author[label1]{Tao Jia}
\ead{tjia@swu.edu.cn}
\address[label1]{College of Computer and Information Science, Southwest University, Chongqing, 400715, P. R. China}

\begin{abstract}
Our ability to control a whole network can be achieved via a small set of driver nodes. While the minimum number of driver nodes needed for control is fixed in a given network, there are multiple choices for the driver node set. A quantity used to investigate this multiplicity is the fraction of redundant nodes in the network, referring to nodes that do not need any external control. Previous work has discovered a bimodality feature characterized by a bifurcation diagram: networks with the same statistical property would stay with equal probability to have a large or small fraction of redundant nodes. Here we find that this feature is rooted in the symmetry of the directed network, where both the degree distribution and the degree correlation can play a role. The in-in and out-out degree correlation will suppress the bifurcation, as networks with such degree correlations are asymmetric under network transpose. The out-in and in-out degree correlation do not change the network symmetry, hence the bimodality feature is preserved. However, the out-in degree correlation will change the critical average degree needed for the bifurcation. Hence by fixing the average degree of networks and tuning out-in degree correlation alone, we can observe a similar bifurcation diagram. We conduct analytical analyses that adequately explain the emergence of bimodality caused by out-in degree correlation. We also propose a quantity, taking both degree distribution and degree correlation into consideration, to predict if a network would be at the upper or lower branch of the bifurcation. As is well known that most real networks are not neutral, our results extend our understandings of the controllability of complex networks.
\end{abstract}


\end{frontmatter}

\section{Introduction}
A dynamical network system is controllable if it can be driven from its initial state to a desired final state by imposing appropriate external signals on a set of its nodes \cite{slotine1991applied,liu2011controllability,kalman1963mathematical}. These nodes, if driven by independent signals, are called driver nodes. With an efficient method, we can identify the minimum set of driver nodes (MDS), through which the whole system can be controlled \cite{liu2011controllability}. The number of driver nodes necessary and sufficient for control, $N_d$, is fixed for a given network but there exist multiple MDSs, and nodes participate in these MDSs differently. Using the likelihood of being in MDS \cite{jia2013control,zhang2019control}, one can classify nodes into three categories \cite{jia2013emergence,zhang2019evolution}: critical nodes must always be controlled and they participate in all MDSs; redundant nodes never require external control hence do not appear in any MDSs; and finally intermittent nodes which act as driver nodes in some but not all control configurations. The fraction of these three categories of nodes can be denoted by $n_c$, $n_r$, and $n_i$, respectively. Naturally, $n_c + n_r + n_i = 1$.

Previous work has discovered a bimodality feature in control \cite{jia2013emergence,jia2014connecting, zhang2019altering, zhang2021altering}. For neutral networks with symmetric out- and in-degree distribution, when the average degree $\langle k\rangle$ of the network exceeds a critical value $k_c$, $n_r$ would follow a bifurcation diagram, which is similar to the pitchfork bifurcation \cite{chen2000bifurcation,wiggins1990introduction}. The bifurcation indicates the existence of two distinct control modes. In one model, corresponding to the upper branch of the bifurcation diagram, most nodes in the network are redundant and only a few nodes can be included in the MDS. In the other model, $n_r$ is on the lower branch of the bifurcation diagram and most nodes can participate in control. On the contrary, for networks with asymmetric out- and in-degree distribution, the bifurcation would be eliminated into a single branch. Specifically, a network whose out-degree distribution is more divergent than its in-degree distribution will stay on the lower branch, and vice versa.

There are intensive studies on the original framework of network controllability \cite{liu2016control,chen2017pinning,xiang2019advances,wang2017physical,liu2017controllability,lou2020towards,tselykh2020influence,wu2020structural,li2019controlling,gates2016control}, with extensions to temporal or multilayer networks \cite{zhang2019evolution,li2017fundamental,posfai2014structural,posfai2016controllability,pang2019controlling}, cost of control \cite{zhang2019evolution,hu2019optimal, sun2018identifying,nie2018control,duan2019energy,sun2017closed,zhao2019research}, and biological and social networks \cite{badyaev2019cycles,ravindran2019network,solimine2020political,yong2020study,liu2020structural,sharma2018controllability,wu2017biomolecular,li2019control,guo2018novel,angulo2019theoretical,yan2017network}. Yet, except for a few studies \cite{posfai2013effect,nie2016effect,takemoto2016analysis,liu2020effect}, most of the previous works mainly consider ``neutral'' networks, whose degree distributions do not have correlations. Here, we extend the analyses to networks with symmetric out- and in-degree distribution under degree correlations. There are 4 types of degree correlation in directed networks \cite{posfai2013effect,foster2010edge}, which affect the bifurcation diagram differently. The in-out degree correlation does not affect $n_r$, leaving the bifurcation diagram intact. The in-in and out-out degree correlation suppress the bifurcation. One branch will vanish and the network only evolves on one strand as the average degree $\langle k\rangle$ increases. The out-in degree correlation will change $k_c$, the critical average degree beyond which the bifurcation occurs. Hence for a network with a fixed average degree, one can tune the out-in degree correlation to make the network sub-critical or super-critical. $n_r$ will follow a bifurcation diagram by changing the out-in degree correlation alone. We propose an analytical approach for $n_r$ in networks with out-in degree correlation, which is approximate but explains how the degree correlation would play a role in the bifurcation of $n_r$. In light of the effect of network symmetry on robust controllability \cite{parastvand2020graph}, and the relationship between the pitchfork bifurcation and system symmetry, we find that the symmetry of the network determines whether bifurcation in control would occur. The bimodality only emerges in symmetric networks and disappears in asymmetric networks, whose asymmetry may be caused by different out- and in-degree distribution or degree correlations. Finally, we propose a quantification of the network symmetry that takes both the degree distribution and degree correlation into consideration, allowing us to make a precise prediction on the control mode of a network. In all, we not only report a new set of numerical observations but also qualitatively explain the rules underlying the patterns observed. Given the fact that most real networks are not neutral, our findings on the correlated networks would bring new insights into the controllability of complex networks.

\section{Results}

\subsection{Problem description}
\begin{figure}[H]
	\begin{center}
		\includegraphics[width=13cm]{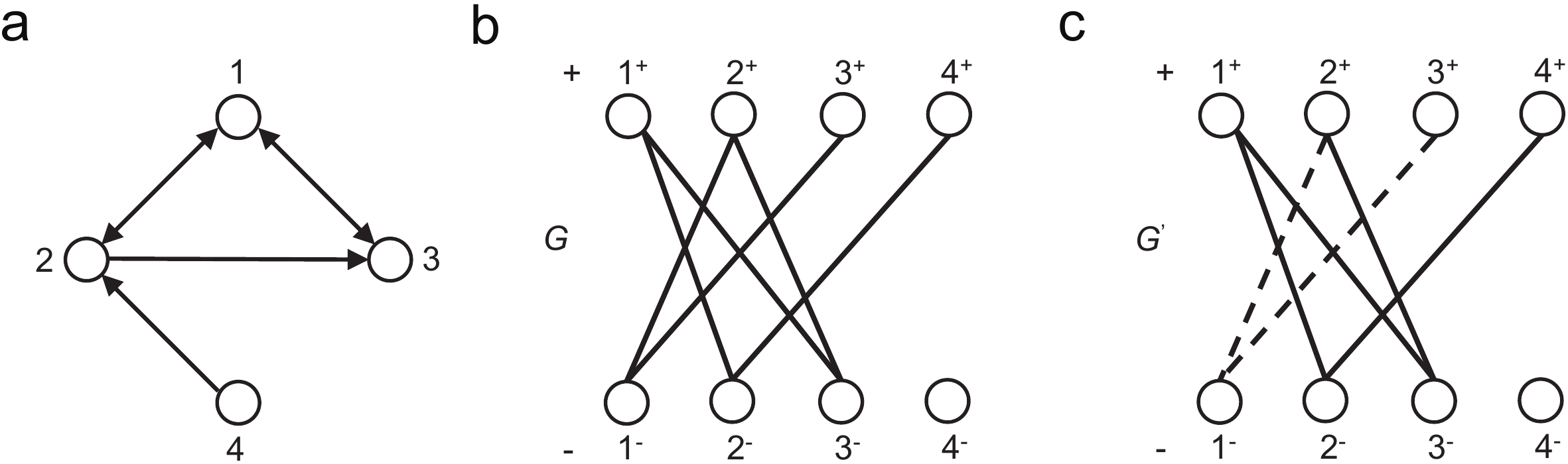}
		\caption{
			({\bf a}) An example of a directed network with four nodes.
			({\bf b}) The bipartite graph represents the directed network in ({\bf a}) where a node \textit{i} is split into two nodes $i^+$ (upper) and $i^-$ (lower). A directed link from node 2 to node 3 in ({\bf a}) corresponds to a connection between node $2^+$ and $3^-$. ({\bf c}) The subgraph $G' = G\backslash i$, obtained by removing node \emph{i} and all its links. According to the theorem, it is obvious that node $1^-$ is always matched in $G$ as its neighbor $3^+$ is not matched in subgraph $G'$.
			\label{fig:fig1}}
	\end{center}
\end{figure}

Consider a directed network $G$ with $N$ nodes and each node has $k_\text{in}$ in-coming links and $k_\text{out}$ out-going links, but without multiply links and self-loop. The directed network can be transformed into a bipartite network by splitting a node $i$ into two nodes $i^{+}$ and $i^{-}$, where node $i^{+}$ records node $i$'s out-going links and node $i^{-}$ records node $i$'s in-coming links (Fig. \ref{fig:fig1}a-b). Consequently, there are two disjoint sets of $+$ and $-$ nodes in the bipartite network. A link only connects a node in $+$ set and a node in $-$ set and there is no link within each set of nodes. To identify the minimum driver nodes necessary and sufficient to control the networked system $G$, one can find a maximum matching in the bipartite network \cite{liu2011controllability,zhao2019controllability}, where a node can at most match another node through one link. If node $j^-$ is unmatched in a maximum matching configuration, node $j$ in $G$ is a driver node for control in this configuration.

One can further identify the role of a node in control, node $j$ for example, by temporally removing node $j^-$ and all its links in the bipartite network and redo the maximum matching \cite{jia2013emergence,zhang2019altering,zhang2021altering}. If node $j^-$ is originally matched and its removal reduces the number of maximally matched nodes, node $j$ is redundant, as node $j^-$ needs to be always matched in the original maximum matching configuration. Correspondingly, node $j$ does not need any external control. It is self-controlled once the sufficient signals are put on other nodes in the network. The fraction of redundant nodes $n_r$ is determined by the degree distribution. It has been found that $n_r$ would follow a bifurcation diagram when the average degree increases \cite{jia2013emergence,zhang2019altering,zhang2021altering}. In this work, we mainly focus on the behavior of $n_r$.

Another quantity we focus on is the degree correlation, which measures the extent that the degrees of two nodes on the ends of a link are correlated. Because a node in the directed network is characterized by both in-degree and out-degree, there are 4 types of degree correlation, \textit{i.e.} the correlation of in-out, in-in, out-out, and out-in degree between the source node and target node (Fig. \ref{fig:fig2}, insert) \cite{posfai2013effect,foster2010edge}. There are several ways to quantify the correlation, like Pearson, Kendall, and Spearman correlation coefficient \cite{bolboaca2006pearson,akoglu2018user}. Due to the ease of calculation, the Pearson correlation coefficient \cite{posfai2013effect,nie2016effect,foster2010edge,newman2002assortative} is frequently applied in previous studies, which is
\begin{equation}
r^{\alpha-\beta} = \frac{L^{-1}\sum_{e}j_{e}^{\alpha}k_{e}^{\beta} - [L^{-1}\sum_{e}1/2(j_{e}^{\alpha} + k_{e}^{\beta})]^{2}}
{L^{-1}\sum_{e}1/2(j_{e}^{\alpha^{2}} + k_{e}^{\beta^{2}}) - [L^{-1}\sum_{e}1/2(j_{e}^{\alpha} + k_{e}^{\beta})]^{2}}\label{(eq.1)},
\end{equation}
where $\alpha, \beta\in\{in, out\}$ represents the type of degree, $j_{e}^{\alpha}$ and $k_{e}^{\beta}$ are the degrees of the source node and target node which are connected by edge \textit{e}, and $L$ is the total number of edges. 

To analyze the effects of degree correlation on the bifurcation diagram of $n_r$, we need to tune the degree correlation. By rewiring links while keeping the initial degree sequences of the neutral network \cite{maslov2002specificity}, we can obtain arbitrary particular degree correlations. Concretely, randomly select two edges \emph{$\widehat{xy}$}, \emph{$\widehat{uv}$} that do not share any node, then swap these two edges to obtain two new edges \emph{$\widehat{xv}$}, \emph{$\widehat{uy}$}. If $d^{\alpha}_x\times d^{\beta}_v + d^{\alpha}_u\times d^{\beta}_y < d^{\alpha}_x\times d^{\beta}_y + d^{\alpha}_u \times d^{\beta}_v$ then correlation $r^{\alpha-\beta}$ decreases, otherwise increases, where $d^{\alpha}_x$, $d^{\beta}_y$, $d^{\alpha}_u$ and $d^{\beta}_v$ are the degrees of the corresponding nodes. Repeat the above procedures until the degree correlation reaches the desired value.

\subsection{The effects of degree correlation on $n_r$}
\begin{figure}[H]
	\begin{center}
		\includegraphics[width=13cm]{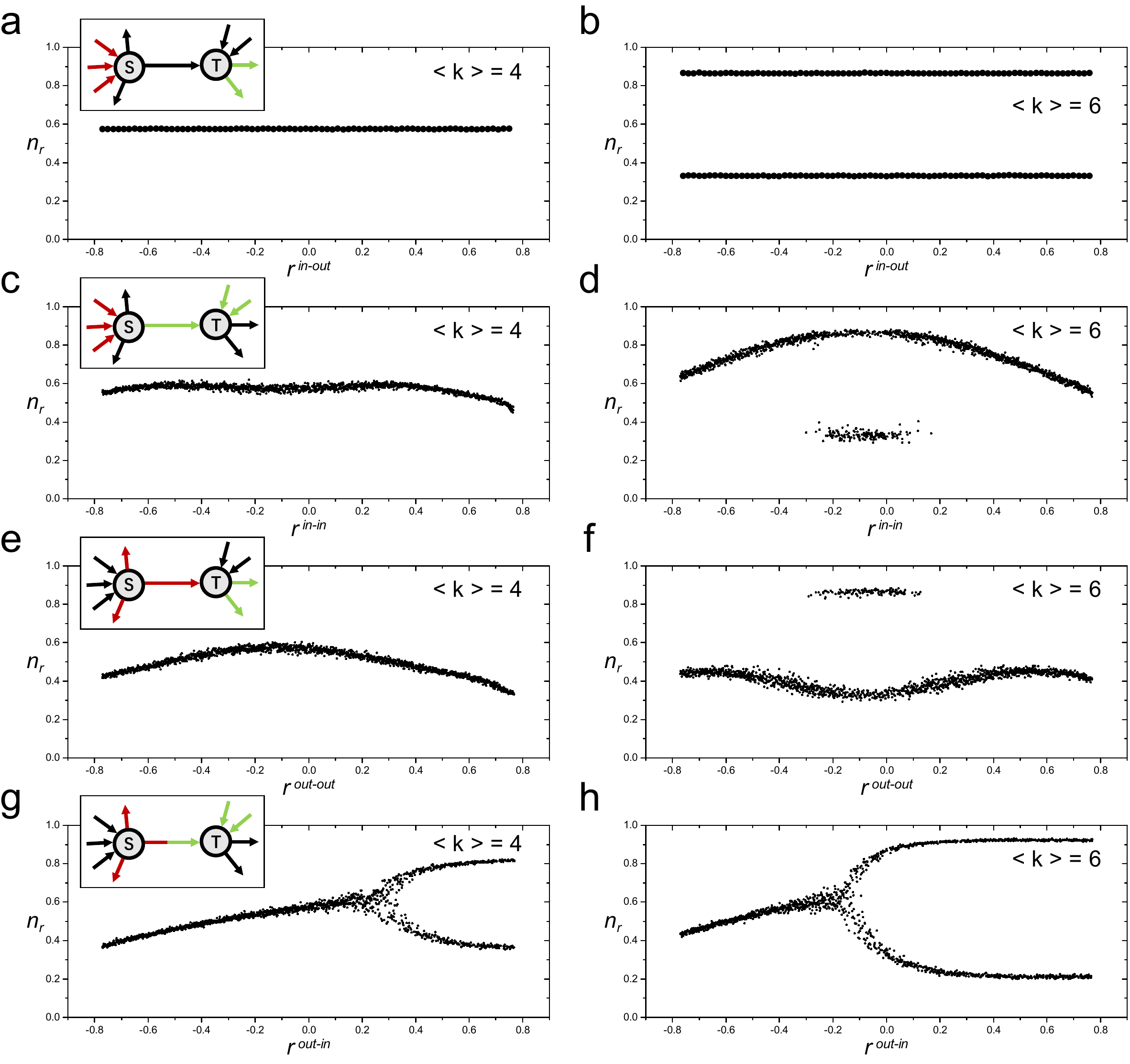}
		\caption{
			We conduct the simulations for sub-critical $\langle k\rangle = 4$ and super-critical $\langle k\rangle = 6$ on Erd\"{o}s-R\'{e}nyi networks with \emph{N} = 10000.
			({\bf a-b}) The results confirmed that in-out correlation has no effect on $n_r$. Specially $n_r$ has two curves when $\langle k\rangle = 6$, as there are two equiprobable values for $n_r$. Each data point is the mean of 100 independent runs.
			({\bf c-d}) For sub-critical $\langle k\rangle = 4$, in-in correlation causes some small perturbations on $n_r$. For super-critical $\langle k\rangle = 6$, in-in correlation suppresses the lower branch of $n_r$ as the magnitude of $r^{in-in}$ increases, regardless of the sign of the correlation strength.
			({\bf e-f}) For sub-critical $\langle k\rangle = 4$, $n_r$ decreases slightly as the magnitude of $r^{out-out}$ increases. For super-critical $\langle k\rangle = 6$, out-out correlation suppresses the upper branch of $n_r$ as the magnitude of $r^{out-out}$ increases, regardless of the sign of the correlation strength.
			({\bf g-h}) For both sub- and super-critical $\langle k\rangle$, out-in correlation induces the bifurcation of $n_r$ when $r^{out-in}$ exceeds a critical value $r^{out-in}_c$, and the critical value $r^{out-in}_c$ decreases as $\langle k\rangle$ increasing.
			\label{fig:fig2}}
	\end{center}
\end{figure}

We fix the average degree $\langle k\rangle$ of a network and vary the degree correlation $r^{\alpha-\beta}$. The four types of degree correlations have different impacts on $n_r$, which can be summarized as follows:

1. The in-out degree correlation does not change $n_r$, whether the network is sub-critical or super-critical (Fig. \ref{fig:fig2}a-b). This result is similar to the previous finding that in-out degree correlation does not alter the fraction of driver nodes $n_d$ \cite{posfai2013effect}.

2. The impact of in-in and out-out degree correlation varies when the network is sub-critical or super-critical. For in-in degree correlation, when $\langle k\rangle < k_c$, the correlation has little impact on $n_r$ (Fig. \ref{fig:fig2}c). When $\langle k\rangle> k_c$, the correlation will eliminate one branch of the bifurcation curve, as $n_r$ will only evolve through the upper branch (Fig. \ref{fig:fig2}d), regardless of the sign of the correlation strength. For out-out degree correlation, when $\langle k\rangle < k_c$, $n_r$ decreases slightly as the magnitude of out-out correlation increases (Fig. \ref{fig:fig2}e). When $\langle k\rangle> k_c$, $n_r$ will only evolve through the lower branch (Fig. \ref{fig:fig2}f), regardless of the sign of the correlation strength. Further, we fix the in-in and out-out degree correlation, and the bifurcation fades away as $\langle k\rangle$ increases (Fig. A.8a-d), which are similar to cases when neutral networks have different out- and in-degree distribution \cite{jia2013emergence}.

\clearpage

3. The out-in degree correlation not only changes the value of $n_r$ but can induce the bifurcation. For a network with $\langle k\rangle < k_c$ or $\langle k\rangle > k_c$, $n_r$ decreases when the correlation becomes more negative. As the correlation becomes more positive, however, a bifurcation occurs when the correlation is over a critical point (Fig. \ref{fig:fig2}g-h). In other words, changing the out-in degree correlation alone can induce the bimodality in control.

\subsection{Out-in correlation induced bimodality}
The bifurcation caused by out-in degree correlation is rooted in the fact that the critical average degree $k_c$ needed for the bifurcation varies with $r^{out-in}$. $k_c$ increases when $r^{out-in}$ becomes more negative, and decreases when $r^{out-in}$ becomes more positive (Fig. A.8e-f). Correspondingly, the bifurcation will occur at a larger $k_c$ when the network has a negative out-in degree correlation. Likewise, by adding more positive out-in degree correlation, a network that is originally sub-critical will become super-critical, displaying the bimodality feature.

To understand how out-in degree correlation changes the evolution of the $n_r$, we need to first briefly review the analytical approach to calculate $n_r$ in the neutral network \cite{jia2013emergence}. As mentioned previously, node $i$ is redundant if the corresponding node $i^-$'s removal will decrease the number of matched nodes. In other words, node $i^-$ should be always matched to get the matching maximized in the bipartite network. Furthermore, it is proved that a node $i$ in a bipartite graph $G$ is not always matched, if and only if all its neighboring nodes are always matched in the subgraph $G'$ where node $i$ and its links are removed. Following the two rules, for a node in the - set whose degree is $k$, the probability that it is not always matched is $(\theta^{+})^k$, where $\theta^{+}$ is the probability that a neighboring node is always matched in the + set of the subgraph $G'$ (Fig. \ref{fig:fig1}b-c). By considering the probability of degree $k$, one can find the fraction of always matched nodes in - set, or equivalently $n_r$ as
\begin{equation}
n_r = 1 - G^{-}(\theta^{+})\label{(eq.2)},
\end{equation}
where $G^{-}(x) = \sum^{\infty}_{k=0}x^{k}P_\text{in}(k)$ is the generating function of degree distribution $P_\text{in}(k)$ \cite{jia2013emergence}.

Let's consider the case when node $j^+$ is the neighbor of a randomly chosen node $i^-$. Eq.(\ref{(eq.2)}) tells that the probability that node $i^-$ is always matched depends on the matching status of $j^+$ in the subgraph $G'$. This would depend on the neighbors of $j^+$, which are in the - set of the bipartite network. Therefore, we can write a similar equation for $\theta^{+}$. The only difference is that node $j^+$ is not a randomly chosen node. Instead, it is a node accessed through a randomly chosen link. Hence, the degree distribution should be replaced by the excess degree distribution, and we get
\begin{equation}
1 - \theta^{+} = \sum^{\infty}_{k=1}Q_\text{out}(k)(\theta^{-})^{k-1} = H^{+}(\theta^{-})\label{(eq.3)}.
\end{equation}
Likewise, if the procedure of the derivation of Eq.(\ref{(eq.3)}) starts at a node $j^-$, we can reach another equation for $\theta^{-}$ by substituting the $+$ sign for $-$ in Eq.(\ref{(eq.3)}) and vice versa, which gives
\begin{equation}
	1 - \theta^{-} = \sum^{\infty}_{k=1}Q_\text{in}(k)(\theta^{+})^{k-1} = H^{-}(\theta^{+}) \label{(eq.4)}.
\end{equation}
$Q_\text{out,in}(k) = P_\text{out,in}(k) \times k/\langle k\rangle$ is the excess degree distribution. $H^{+}(x)$ and $H^{-}(x)$ are the generating functions of excess degree distribution $Q_\text{out}$ and $Q_\text{in}$, respectively. The power index $k-1$ refers to the fact that one link connects to $i^-$ is removed in the subgraph $G'$.

Combining Eqs.(\ref{(eq.3)}) and (\ref{(eq.4)}) eventually gives an equation for $\theta^{+}$ as
\begin{equation}
1 - \theta^{+} = H^{+}(1 - H^{-}(\theta^{+}))\label{(eq.5)},
\end{equation}
The numerical solution of Eq.(\ref{(eq.5)}) provides the value of parameter $\theta^{+}$ in Eq.(\ref{(eq.2)}), and the number of solutions determines if there is a bifurcation of $n_r$.

Eqs.(\ref{(eq.2)}) to (\ref{(eq.5)}) work in neutral networks only, in which degree correlation is absent. However, because the out-in degree correlation in the directed network directly maps to the degree correlation of a bipartite network, Eqs.(\ref{(eq.2)}) to (\ref{(eq.5)}) can serve as fundamentals to understand systems with correlations. Assume now that the bipartite network has a degree correlation $r$, which is equivalent to the out-in degree correlation $r^{out-in}$ in the directed network. Note that the conclusion still holds that a node $i^-$ is not always matched if and only if all its neighbors are always matched in $G'$, which gives
\begin{equation}
n_r = 1 - \sum_{k}^{\infty}P_\text{in}(k)(\theta^{+}(r,k))^k \label{(eq.6)}.
\end{equation}

Because the network is not neutral, the probability that a neighboring node is always matched in $G'$ would depend on its degree $k$ and the degree correlation $r$, making the sum in Eq.(\ref{(eq.6)}) intractable. Note that the sum over $k$ eventually eliminates the dependence of $k$. Hence we can consider a mean-field approximation that
\begin{equation}
\sum_{k}^{\infty}P_\text{in}(k)(\theta^{+}(r,k))^k \approx \sum_{k}^{\infty}P_\text{in}(k)(\theta^{+}_o + \alpha r)^k = G^{-}(\theta^{+}_o + \alpha r), \label{(eq.7)}
\end{equation}
where $\theta^{+}_o$ is the mean value and the term $\alpha r$ captures the effect of degree correlation.

\begin{figure}[H]
	\begin{center}
		\includegraphics[width=13cm]{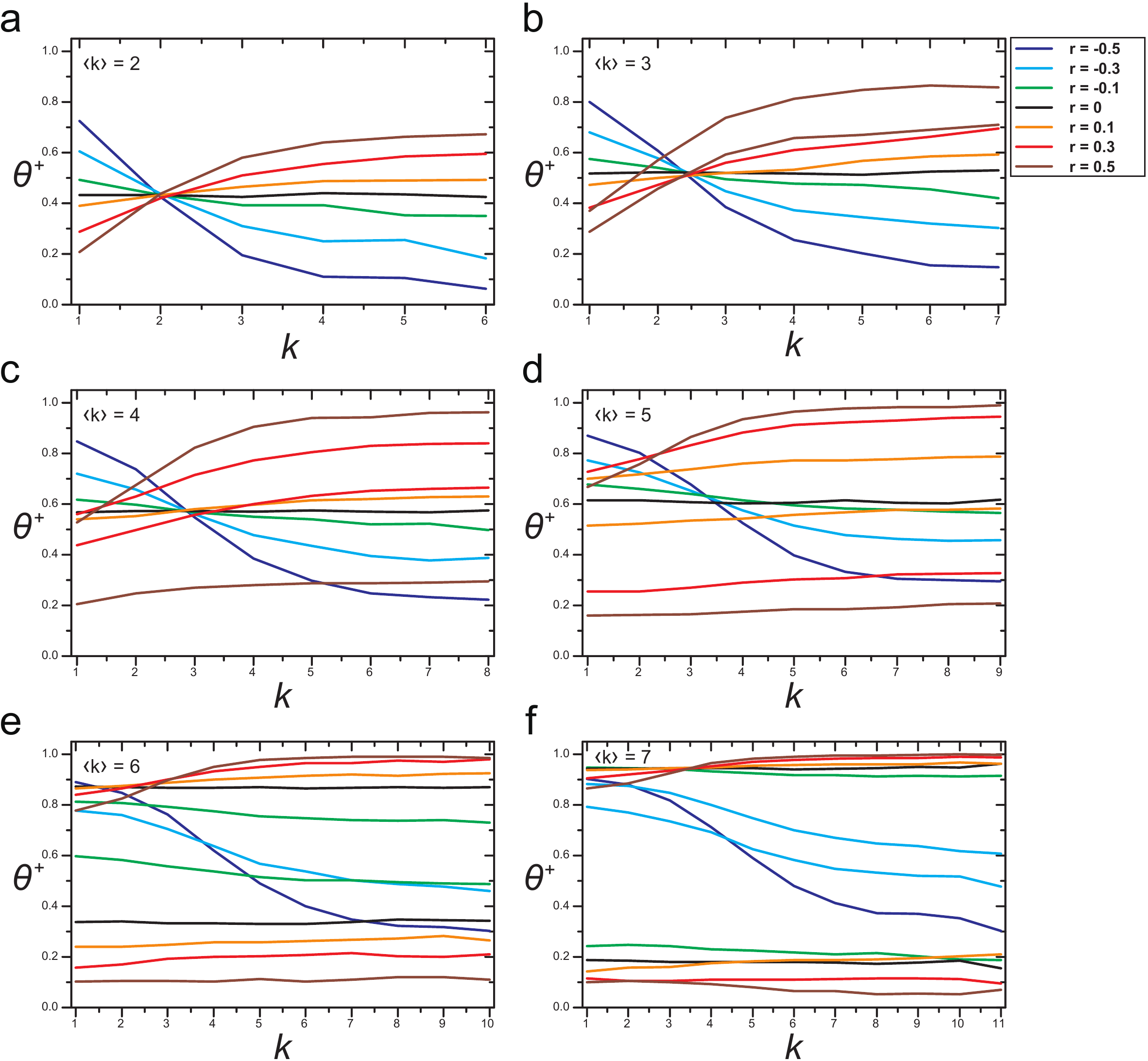}
		\caption{
			Here we fix $r = 0, \pm0.1, \pm0.3, \pm0.5$. For different $\left\langle k\right\rangle $, a positive \textit{r} tends to induce a positive correlation between $\theta^{+}$ and $k$, and this correlation gets stronger as \textit{r} increases. While the denser the network is, the weaker the correlation is. The impact of negative \textit{r} is similar but inducing a negative correlation between $\theta^{+}$ and $k$. That's to say, the value of perturbation factor $\alpha r \sim r$, which confirms that $\alpha$ in Eq.(\ref{(eq.7)}) is positive.
			\label{fig:fig3}}
	\end{center}
\end{figure}

It is important to first determine the sign of $\alpha$ in Eq.(\ref{(eq.7)}) as it controls how the positive and negative degree correlations play the role. $\theta^{+}$ is the probability that a neighboring node is always matched in the + set of the subgraph $G'$. By definition, we numerically measure $\theta^{+}$ by calculating the fraction of always matched neighbors of a node in - set with degree $k$. The simulation results demonstrate that when the network is neutral, $\theta^{+}$ is the same for all $k$, in line with our intuition. When $r > 0$, $\theta^{+}$ increases with $k$, demonstrating a higher value on average (Fig. \ref{fig:fig3}). On the contrary, when $r < 0$, $\theta^{+}$ decreases with $k$, making the overall level of $\theta^{+}$ lower. Hence, it is reasonable to conclude that $\alpha$ in Eq.(\ref{(eq.7)}) is positive.

Using the approximation by Eq.(\ref{(eq.7)}), we can rewrite Eqs. (\ref{(eq.3)}) and (\ref{(eq.4)}) as
\begin{equation}
	1 - \theta^{+}_{o} = H^{+}(\theta^{-}_{o} + \alpha r)\label{(eq.8)},
\end{equation}
\begin{equation}
	1 - \theta^{-}_{o} = H^{-}(\theta^{+}_{o} + \alpha r)\label{(eq.9)}.
\end{equation}

\begin{figure}[H]
	\begin{center}
		\includegraphics[width=13cm]{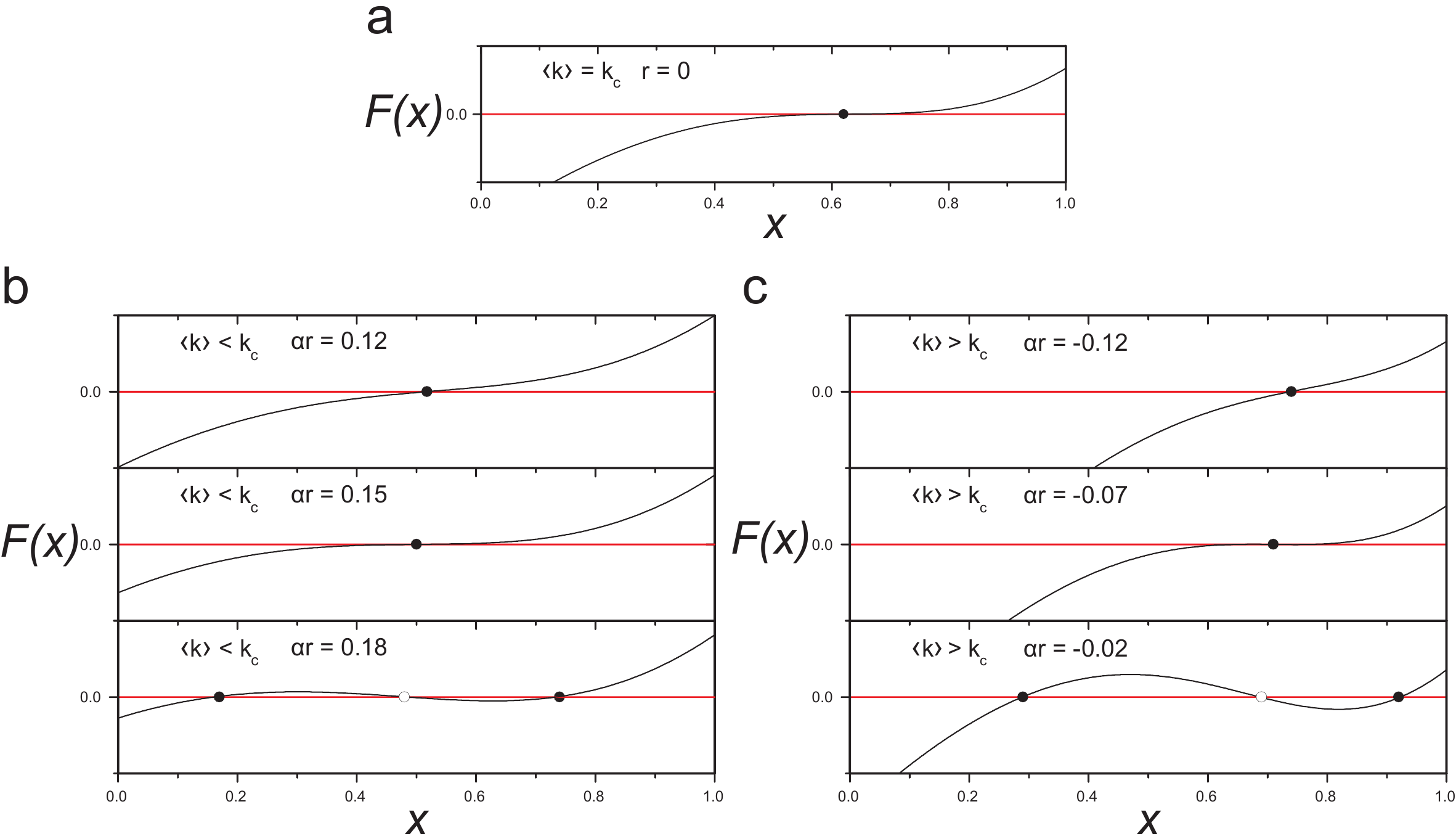}
		\caption{The curves $F(x) = H^{+}(1 + \alpha r - H^{-}(x + \alpha r)) + x - 1$ are numerical simulations of Eq.(\ref{(eq.11)}).
			({\bf a}) In the neutral case for $r = 0$ and $\langle k\rangle = k_c$, the curve of $F(x)$ is tangent to $y = 0$.
			({\bf b}) For a random sparse network whose average degree $\langle k\rangle < k_c$, when $r$ is positive and small, there is one solution for equation $F(x) = 0$ corresponding to that $n_r$ monotonically increases. When $r$ excesses a critical value the curve of $F(x)$ is tangent to $y = 0$, which means there will be a positive $r$ inducing the bifurcation for a sparse network with $\langle k\rangle < k_c$. As $r$ increase, equation $F(x) = 0$ develops to three solutions but only two of which are stable, corresponding to two branches of $P(n_r)$.
			({\bf c}) The same goes for denser networks with $\langle k\rangle > k_c$ and negative $r$. The conditions of sub-critical, critical, and super-critic are met when the curve of $F(x)$ intersects at one point, tangentially, and at multiple points with $y = 0$, respectively.
			\label{fig:fig4}}
	\end{center}
\end{figure}

Combining Eqs.(\ref{(eq.8)}) and (\ref{(eq.9)}), we obtain
\begin{equation}
1 - \theta^{+}_{o} = H^{+}(1 +\alpha r - H^{-}(\theta^{+}_{o} + \alpha r))\label{(eq.10)}.
\end{equation} 

Consequently, the bifurcation depends on the number of solutions of
\begin{equation}
F(x) = H^{+}(1 +\alpha r - H^{-}(x + \alpha r)) + x - 1 = 0\label{(eq.11)}.
\end{equation}

We admit that Eq.(\ref{(eq.11)}) may appear to be an oversimplification of the essential problem. Indeed, when the degree correlation presents, the original analytical framework may not be valid anymore (Fig. \ref{fig:fig4}a). Nevertheless, our approximation gives the correct description of how degree correlation $r$ would affect $n_r$. By increasing $r$, one can tune the curve of $F(x)$ and make Eq.(\ref{(eq.11)}) go from 1 solution to 2 (Fig. \ref{fig:fig4}b). Correspondingly, the curve of $n_r$ changes from sub-critical to super-critical. Likewise, by bringing more negative correlation, we can change the system that is originally super-critical to sub-critical, where Eq.(\ref{(eq.11)}) can generate only 1 solution (Fig. \ref{fig:fig4}c). This agrees well with our observation that a bifurcation could occur if we fix $\langle k\rangle$ of the network and tune $r^{out-in}$ alone (Fig. \ref{fig:fig2}g-h).

\subsection{The degree symmetry of network explains the emergence and disappearance of bifurcation}
The numerical observation on the emergence and disappearance of bifurcation prompts us to ask: what physical quantity is associated with this phenomenon. Note that even in neutral networks, the difference between the out- and in-degree distribution can also make one branch of the bifurcation disappear. Hence it is crucial that the quantity can unify both observations from the degree correlation and the difference of degree distribution. 

Because the bifurcation diagram in control is similar to the pitchfork bifurcation in continuous dynamical systems, which occurs generically in physical systems that possess an underlying symmetry \cite{wiggins1990introduction,marques2013fold}. Meanwhile, imperfect symmetry or asymmetry will make the pitchfork bifurcation unstable \cite{golubitsky1979imperfect,crawford1991symmetry}. Motivated by this pattern, we conclude that bifurcation can only emerge in symmetric networks. Here, we define the ``symmetry'' following the concept in physics that a symmetry of a physical system is a physical or mathematical feature of the system (observed or intrinsic) that is preserved or remains unchanged under some transformation \cite{gross1996role,brading2003symmetries}. The transformation we consider is the network transpose, \textit{i.e.} flipping the direction of all links in the network \cite{badyaev2019cycles, gao2016universal}. The symmetry thus gauges if the same statistical property of the network topology holds after the transposition. For example, a neutral network with identical out- and in-degree distribution is symmetric, because the statistical property of the network remains unchanged under transposition. A neutral network whose out- and in-degree distributions are different is asymmetric, as the in-degree distribution of the original network is different from that of the transposed network. 

\begin{figure}[H]
	\begin{center}
		\includegraphics[width=13cm]{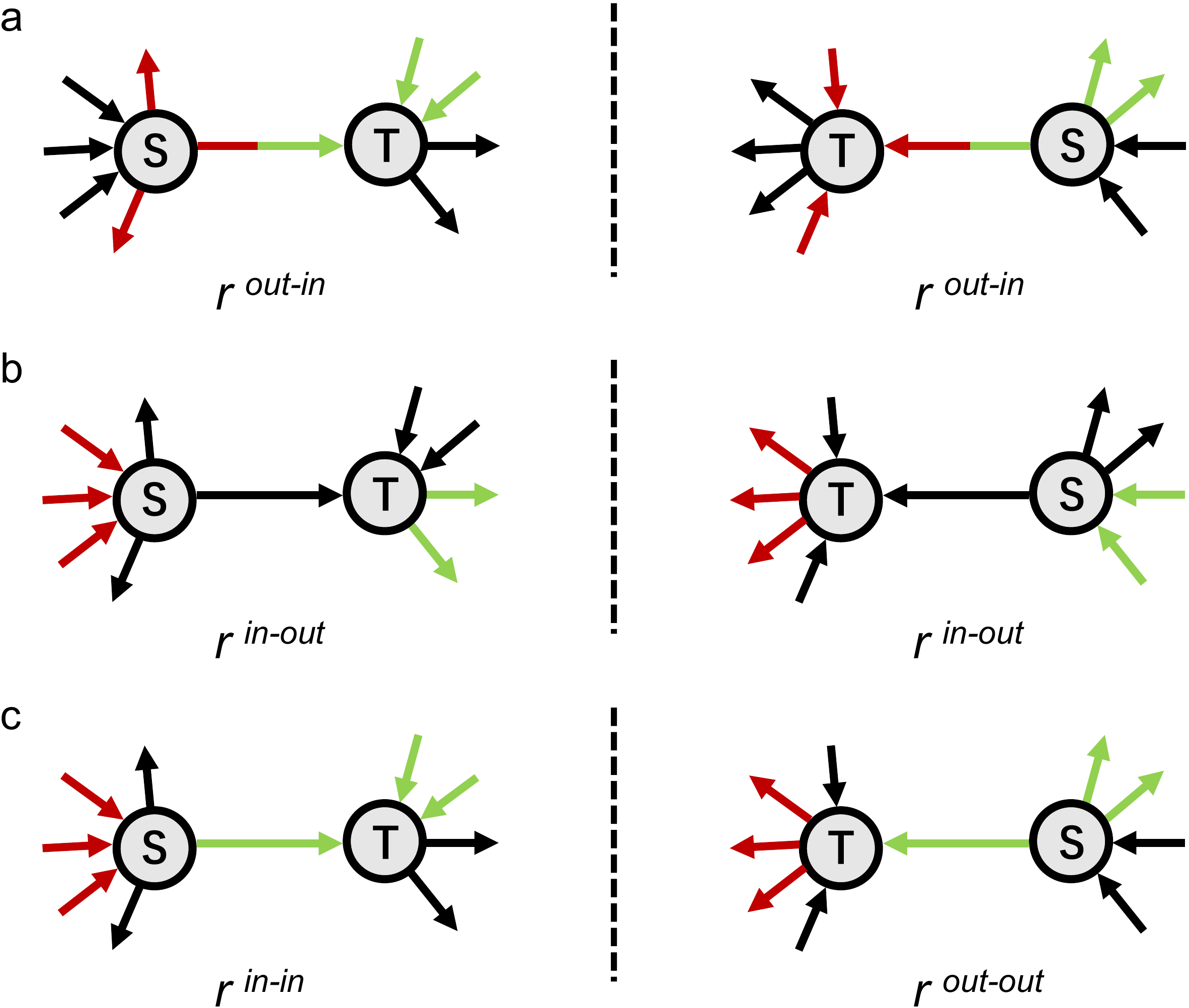}
		\caption{ Examples for network transpose, where $S$ and $T$ represent source node and target node respectively.
			({\bf a-b}) In the original network and the transposed network, the out-in and in-out degree correlation are the correlations between the colored links. They remain unchanged under the network transpose. Therefore, the network does not show any different statistical property under the network transpose. The out-in and in-out degree correlation does not alter the network symmetry.
			({\bf c}) A network with in-in degree correlation would demonstrate an out-out degree correlation under the network transpose. There exist different statistical properties between the original and transposed network. Thus, the in-in degree correlation changes the symmetry of the network.
			\label{fig:fig5}}
	\end{center}
\end{figure}

The definition of symmetry can be extended to cases where degree correlation exists. Consider a network with an out-in degree correlation, characterizing the fact that the out-degree of the source node and the in-degree of the target node connected by a link are not independent. After the transposition, the in-degree becomes the out-degree in the transposed network, and vice versa. But two nodes originally connected by a link remain unchanged and so does the dependency in their out- and in-degree (Fig. \ref{fig:fig5}a). Therefore, the out-in degree correlation preserves in the transposed network to the same extent as the original network. If the out- and in-degree distribution are the same, the network is symmetric. The same conclusion holds for the in-out degree correlation (Fig. \ref{fig:fig5}b). On the contrary, a network with an in-in degree correlation is asymmetric (Fig. \ref{fig:fig5}c). This is because the degree correlation in the transposed network becomes out-out. The transposed network demonstrates a new property that the original network does not contain. 

With the definition of network symmetry and numerical results, we can see that bifurcation only occurs in symmetric networks (Fig. A.8e-f). In asymmetric networks, the evolution of $n_r$ would only follow one curve (Fig. A.8a-d). By comparing $n_r$ of the original network with that of the transposed network, we can even tell if a network is on the upper or lower branch. A question then arises: what quantity can be applied to predict if a network would have a larger or smaller $n_r$. The question has been partially answered in neutral networks \cite{jia2013emergence}, where the difference between the variance of the out- and in-degree distribution can predict if a network is on the upper or the lower branch of the bifurcation curve. However, previous results do not apply in networks with degree correlation, since the variance of degree distribution does not change under degree correlation. The measure of degree correlation itself can be used to predict if $n_r$ is on the upper or lower branch, but it alone is insufficient because degree correlation is zero in neutral networks with different out- and in-degree distribution. Hence, a more comprehensive measurement, which takes both the degree correlation and difference between out- and in-degree distribution into consideration, is needed.

Here we introduce the new measure using the conditional degree distribution $P(k^{'}|k)$, defined as the probability that a node with degree $k$ has a neighbor node with degree $k^{'}$ \cite{newman2001random,schwartz2002percolation,hooyberghs2010biased}. In neutral networks, $P(k^{'}|k)$ is independent of $k$. Therefore $P(k^{'}|k) = k^{'}P(k^{'})/\langle k\rangle$ equals to the excess degree distribution $Q(k^{'})$. In networks with degree correlation, $P(k^{'}|k)$ depends on $k$ providing a measure of the extent of correlation \cite{newman2002assortative,vazquez2002large}. In other words, if the neighbors of a node with degree $k$ tend to have similar degree $k^{'}$, the conditional degree distribution $P(k^{'}|k)$ would be more centralized than in neutral case, and vice versa. 

We consider the ratio between the square root of the second moment and the first moment of $P(k^{'}|i)$ as the measure of the degree divergence of node $i$'s neighbors
\begin{equation}
D^{i} = \frac{\sqrt{\sum_{k^{'}}k^{'2}*P(k^{'} | i)}}{\langle k^{'}\rangle}\label{(eq.12)},
\end{equation}
where $\langle k^{'}\rangle = k_{nn}(i) = \sum_{k^{'}}k^{'}*P(k^{'} | i)$ is the average nearest neighbor degree of node $i$. In directed networks, we need to consider the out- and in-degree separately. The out link of the node $i$ is the incoming link of its neighbors, and vice versa, which gives
\begin{equation}
D^{i}_\text{out,in} = \frac{\sqrt{\sum_{k_\text{in,out}^{'}}k_\text{in,out}^{'2}*P(k_\text{in,out}^{'} | i_\text{out,in})}}{\langle k_\text{in,out}^{'}\rangle}\label{(eq.13)}.
\end{equation}
To comprehensively quantify the effect of degree correlations and degree distributions on network’s statistical properties, we normalize the average neighbor degree divergence by the inherent degree divergence, which defines the symmetry coefficient $S_\text{out}$ and $S_\text{in}$ for both out- and in-degree as
\begin{equation}
S_\text{out,in} = \frac{\overline{D^{i}_\text{out,in}}}{\langle k_\text{in,out}^2\rangle}\label{(eq.14)}.
\end{equation}

We propose that the ratio $S_\text{out}/S_\text{in}$ can be used to predict the symmetry of networks since a symmetric network must meet $S_\text{out}=S_\text{in}$. Otherwise, the network will show a different statistical property under the transpose, indicating that the network is asymmetric.

\begin{figure}[H]
	\begin{center}
		\includegraphics[width=13cm]{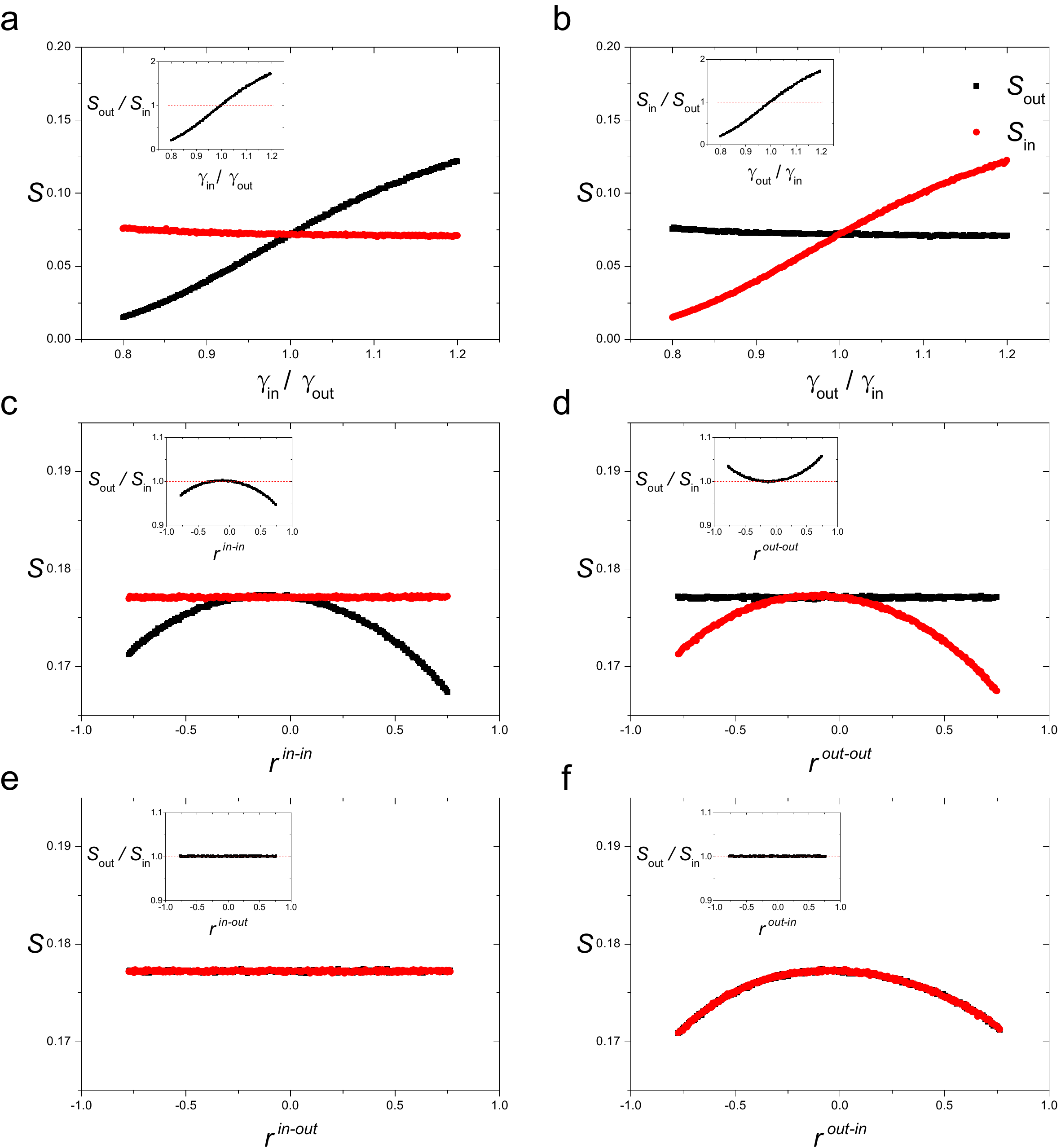}
		\caption{
			({\bf a-b}) For neutral Scale-Free networks with $\langle k\rangle = 4$, we change the symmetry between the out- and in-degree distribution by fixing $\gamma_\text{out}$($\gamma_\text{in}$)=2.5 and varying $\gamma_\text{in}$($\gamma_\text{out}$). When $\gamma_\text{out} > \gamma_\text{in}$, $S_\text{out}<S_\text{in}$, making $n_r$ follow the upper branch of the bifurcation. On the contrary, when $\gamma_\text{out} < \gamma_\text{in}$, $S_\text{out} > S_\text{in}$, leading $n_r$ to follow the lower branch of the bifurcation. There is a good matching between the analytical and numerical results of the ratio $S_\text{out}/S_\text{in}$ (insert).
			({\bf c-d}) The in-in degree correlation induces $S_\text{out} < S_\text{in}$ and the out-out degree correlation induces $S_\text{out} > S_\text{in}$.
			({\bf e-f}) The in-out and out-in degree correlation does not change $S_\text{out} / S_\text{in}$. Although the actual value of $S_\text{out}$ and $S_\text{in}$ are altered when the out-in degree correlation presents, the ratio remains to be 1. Each data point is the mean of 100 independent runs. 
			\label{fig:fig6}}
	\end{center}
\end{figure}

For neutral networks, $S_\text{out}/S_\text{in}$ can be explicitly calculated as
\begin{equation}
	S_\text{out}/S_\text{in} = \sqrt{\frac{\langle k_\text{in}^3\rangle}{\langle k_\text{out}^3\rangle}} \times
	\frac{\langle k_\text{out}^2\rangle^2}{\langle k_\text{in}^2\rangle^2}\label{(eq.15)}.
\end{equation}
This indicator is very similar to that in previous work \cite{jia2013emergence}. The only difference is a correction term based on the third moments of the degree distributions. But overall, $S_\text{out}/S_\text{in}$ depends mostly on the second moments. The results are further verified by numerical simulations (Fig. \ref{fig:fig6}a-b). When $S_\text{out}/S_\text{in} < 1$, $n_r$ follows the upper branch. When $S_\text{out}/S_\text{in} > 1$, $n_r$ follows the lower branch.

For networks with degree correlations, $S_\text{out}/S_\text{in}$ can be numerically quantified and the results are summarized as follows:

1. \textit{in-in correlation.} In-in correlation induces $S_\text{out} < S_\text{in}$ (Fig. \ref{fig:fig6}c). Correspondingly, $n_r$ follows the upper branch (Fig. A.8a-b).

2. \textit{out-out correlation.} We obtain out-out correlation by switching the direction of each link of in-in correlation without changing the matching of networks. So $S_\text{out}/S_\text{in}$ and $P(n_r)$ can be obtained by exchanging in-degree sequence and out-degree sequence. $S_\text{out} > S_\text{in}$ (Fig. \ref{fig:fig6}d) gives rise to the lower branch of $n_r$ (Fig. A.8c-d);

3. \textit{in-out correlation.} In-out correlation does not change $S_\text{out}$ or $S_\text{in}$. $S_\text{out} = S_\text{in}$ as long as the out- and in-degree distribution are the same (Fig. \ref{fig:fig6}e). This is in line with our numerical findings that the in-out degree correlation does not change $n_r$ (Fig. \ref{fig:fig2}a-b).

4. \textit{out-in correlation.} Out-in correlation has no effect on network's symmetry neither (Fig. \ref{fig:fig6}f, insert). Therefore, bifurcation would emerge in networks with out-in correlation. Although the critical point is altered by the degree correlation.

\begin{figure}[H]
	\begin{center}
		\includegraphics[width=11.3cm]{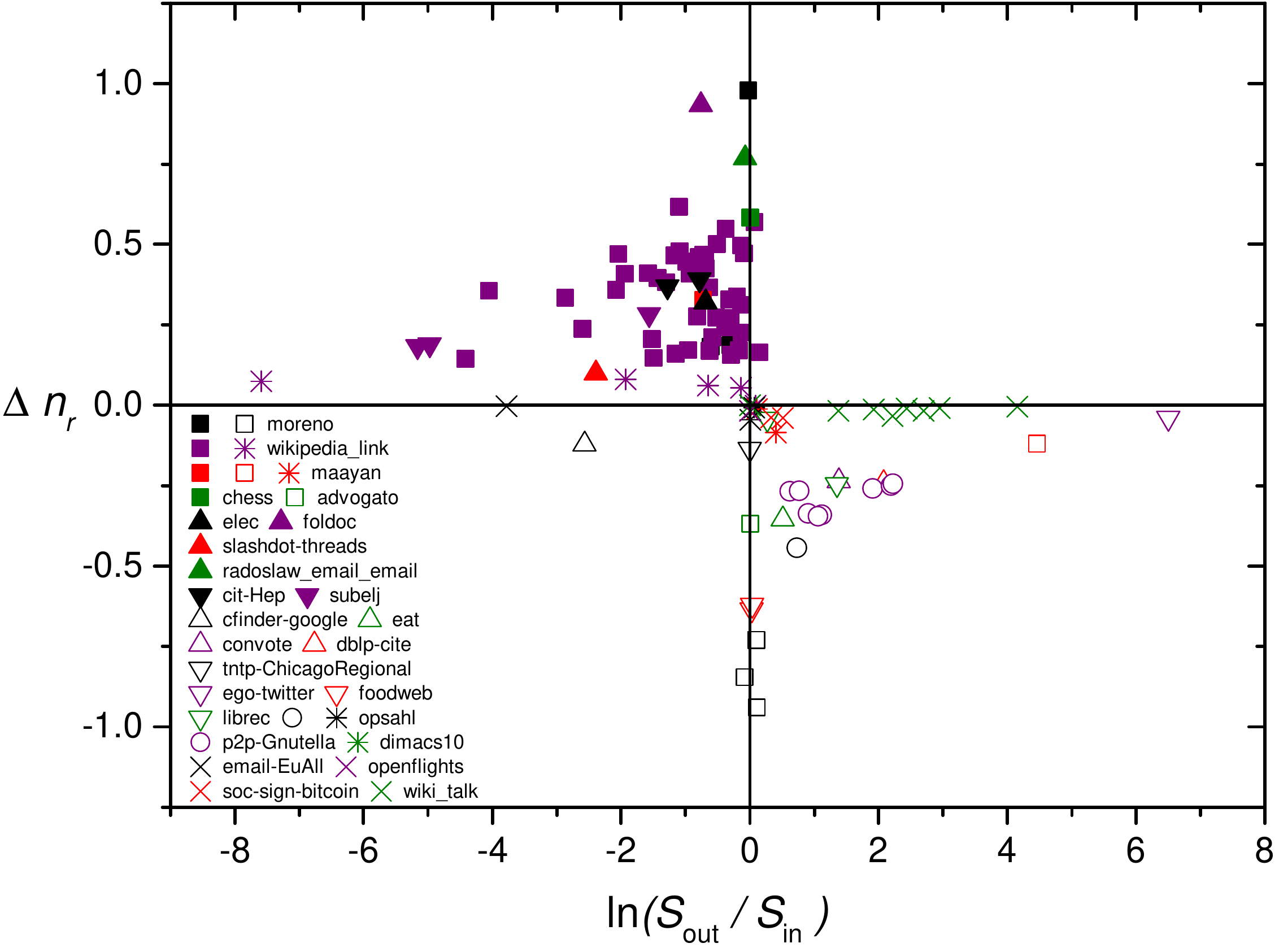}
		\caption{
			The plot of $\Delta n_r$ and ln$(S_\text{out}/S_\text{in})$ for real networks. Most points fall in regions $(x\geq 0,y\leq0)$ and $(x\leq 0,y\geq 0)$, indicating that the difference between symmetry coefficient $S_\text{out}$ and $S_\text{in}$ of the out- and in-degree sequence determines the mode of $n_r$. Solid symbols represent networks with $\Delta n_r > 0$ (centralized control mode in upper branch), hollow symbols correspond to $\Delta n_r < 0$ (distributed control mode in lower branch), and star symbols are for $\Delta n_r \approx 0$. The networks we analyzed here are available at http://konect.cc/.
			\label{fig:fig7}}
	\end{center}
\end{figure}

To bring more supports to our proposed measure, we calculate the $S_\text{out}/S_\text{in}$ of 105 real networks. We also measure the $n_r$ of the original network and its transposed networks (denoted by $n^T_r$). If $n_r$ of the original network follows the upper branch, then it would follow the lower branch in the transposed network. Hence, $\Delta n_r=n_r-n^T_r$ can be used to gauge if the network is on the upper or lower branch of the bifurcation \cite{jia2013emergence,zhang2019altering,zhang2021altering}. The prediction we made, which is also supported by numerical results of model networks, is that $\Delta n_r > 0$ when $S_\text{out}/S_\text{in} < 1$ and $\Delta n_r < 0$ when $S_\text{out}/S_\text{in} > 1$ (Fig. \ref{fig:fig7}). Indeed, despite the variety of degree distribution and degree correlation in real networks, we find they roughly all fall into the second and fourth quadrant, which is in very nice agreement with our prediction.

\section{Discussion}
To summarize, we study the effects of degree correlations on the fraction of redundant nodes ($n_r$) in network control. We find that the four types of different degree correlations bring different influences. The in-out degree correlation does not change $n_r$. The in-in and out-out degree correlations slightly change the magnitudes of $n_r$. But more importantly, they eliminate the bimodality feature of $n_r$, making it evolve only through one branch of the bifurcation curve. The out-in correlation changes the critical value of $k_c$ when the bifurcation occurs. Hence tuning the out-in correlation alone can induce a bifurcation when the average degree of the network is fixed. 

To bring more insights on how the out-in correlation affects the evolution of $n_r$, we apply the mean-field approximation on the original analytical framework applied in neutral networks. As an approximation, this approach adequately demonstrates how the degree correlation can make a system from sub-critical to super-critical. We further show that the emergence and disappearance of the bifurcation can be explained by the symmetry in the directed network, which can be affected by both the degree correlation and the difference in the out- and in-degree distribution. Using the measure we proposed to quantify network symmetry, we find a nice match in both model and real networks. 

We admit that some results are preliminary and worth further investigation. In neutral networks, we can analytically show how the difference in the out- and in-degree distribution plays a role in the equation of $n_r$. This part is currently unknown due to the difficulties in the analytical calculation involving degree correlations. Even our approximate deduction with the out-in correlation is just the first step towards the full understanding of the observations. It would be also interesting to analyze how the degree correlation would change an individual node's role in control. Besides, most results in our paper are derived qualitatively. This is partially because the answer to the question we ask is binary. In Section 2.2, we are interested if a bifurcation could occur or not under degree correlation. In Section 2.4, we are investigating if a network is on the upper branch or the lower. To answer these questions, we do not necessarily need an exact equation to match the numerical simulation. More particularly, what we need in section 2.4 is an indicator to classify the network into one of the two categories (upper or lower branch). For that purpose, a qualitative measure is sufficient. Nevertheless, a quantitative description of the system is important and worth investigating, which is, on the other hand, challenging. We are not able to perform it at this stage. But the phenomenon we discover and the network symmetry we propose can arouse future research on this direction. The relationship between the network symmetry and degree correlation would be useful in the study of controllability in real networked systems. These systems are usually characterized by degree correlations and non-identical out- and in-degree distribution. All these questions can be addressed in future studies.

\section*{Acknowledgement}
This work is supported by the Natural Science Foundation of China (No. 61603309 and No. 62006198).

\appendix
\section{}
\begin{figure}[H]
	\begin{center}
		\includegraphics[width=13cm]{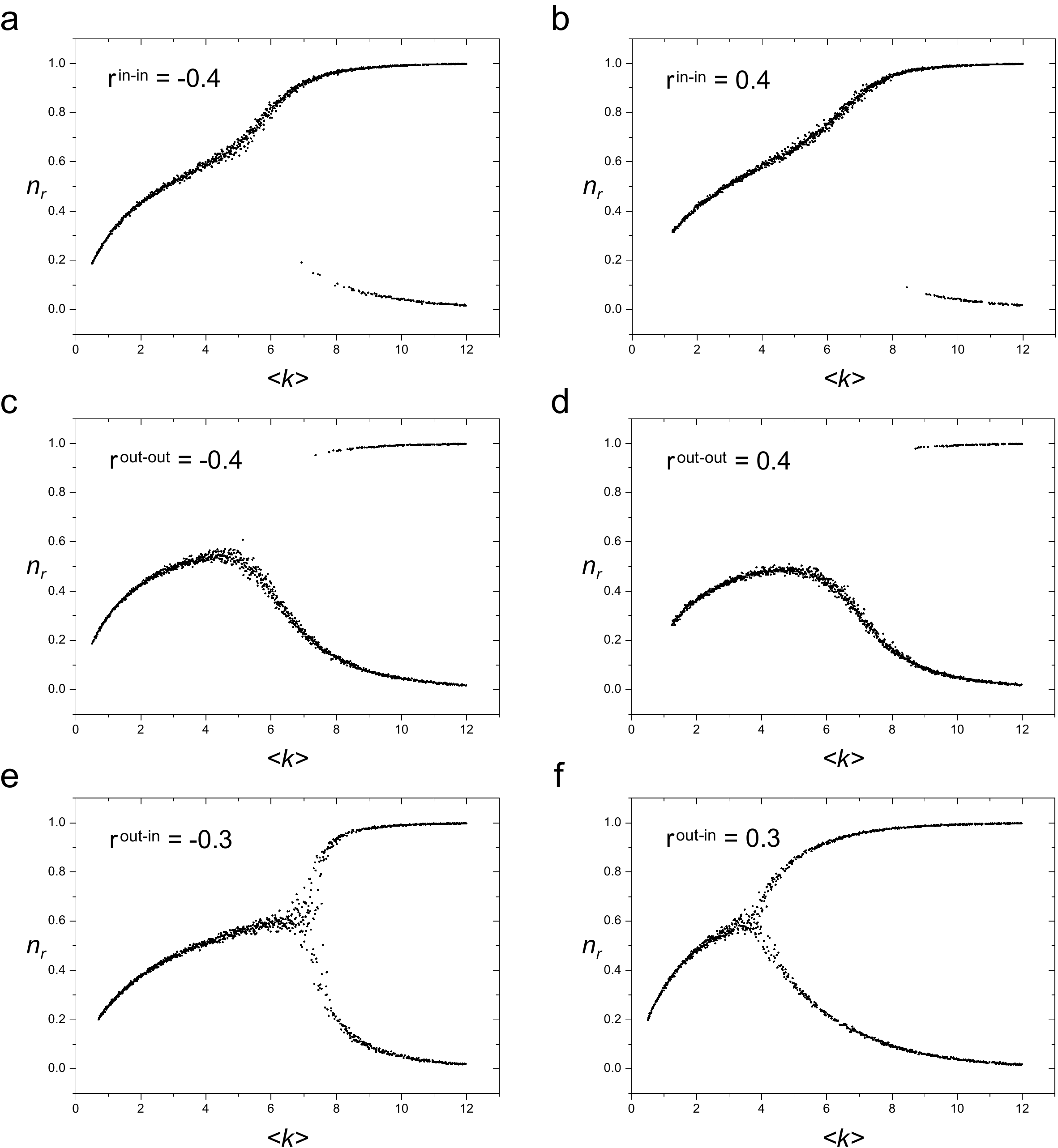}
		\caption*{Figure A.8:
			Fixing $r^{\alpha-\beta}$ and varying $\langle k\rangle$.
			({\bf a-b}) in-in degree correlation leads to the upper branch, regardless of the sign of the correlation strength. 
			({\bf c-d}) out-out degree correlation leads to the lower branch, regardless of the sign of the correlation strength.
			According to Fig. \ref{fig:fig2}d and Fig. \ref{fig:fig2}f, the suppression becomes more obvious as the value of $r^{in-in}$ and $r^{out-out}$ increases.
			({\bf e-f}) out-in degree correlation keeps the bifurcation and changes the $k_c$, as the more positive $r^{out-in}$ is, the smaller $k_c$ is.
			\label{fig:figA1}}
	\end{center}
\end{figure}
\clearpage

\bibliographystyle{elsarticle-num} 
\bibliography{arxiv}

\end{document}